\documentclass[twocolumn,amsmath,amssymb]{revtex4}

\usepackage{amsmath}
\usepackage{graphicx}

\begin{document}
\title{Charge transport through graphene junctions with wetting metal leads}
\author{Salvador Barraza-Lopez}
\email{sbarraza@uark.edu}
\affiliation{School of Physics, Georgia Institute of Technology,
Atlanta, Georgia 30332, USA}
\affiliation{Department of Physics, University of Arkansas,
Fayetteville, Arkansas 72701, USA}
\author{Markus Kindermann}
\affiliation{School of Physics, Georgia Institute of Technology,
Atlanta, Georgia 30332, USA}
\author{M. Y. Chou}
\email{meiyin.chou@gatech.edu}
\affiliation{School of Physics, Georgia Institute of Technology,
Atlanta, Georgia 30332, USA}
\affiliation{Institute of Atomic and Molecular Sciences, Academia Sinica, Taipei 10617, Taiwan}

\begin{abstract}
Graphene is believed to be an excellent candidate material for next-generation electronic devices.
However, one needs to take into account the nontrivial effect of metal contacts in order to precisely control the charge injection and extraction processes. We have performed transport calculations for graphene junctions with wetting metal leads (metal leads that bind covalently to graphene) using nonequilibrium Green's functions and density functional theory. Quantitative information is provided on the increased resistance with respect to ideal contacts and on the statistics of current fluctuations.  We find that charge transport through the studied two-terminal
graphene junction with Ti contacts is pseudo-diffusive up to surprisingly high energies.
\end{abstract}
\date{Published on June 8, 2012}


\maketitle

Metal leads play a crucial role in the transport of charges through small-scale graphene junctions \cite{Leonard,Lee,GeimSSC,Huard,Japan2010,Venugopal,APL2011,Avouris2011,Chris}, and understanding the phenomena induced at these leads has significant technological implications.  A basic, yet relevant question becomes: How should one model real metal contacts on graphene devices? Previous theoretical modeling of charge transport through graphene junctions with metal contacts has made two general assumptions \cite{Huard,Beenakker,Cayssol,Golizadeh,Dolfus,BarrazaPRL2010,Venugopal,Nouchi,Avouris2011,Titov}:
that there exists a simple charge transfer and a resulting band bending $\Delta$ at the leads \cite{Kelly1,Kelly2,Cayssol,BarrazaPRL2010,Khomyakov2010,Lee,Huard,PRBArpesGrapheneMetal2010},
and that the electronic dispersion at the leads is linear and graphene-like.
In addition, level broadening was added to some of the models to further increase the charge density at the lead \cite{Golizadeh,Dolfus,BarrazaPRL2010,Avouris2011,Titov}.
These approaches are inadequate to describe   wetting metal/graphene interfaces such as Ti/graphene \cite{Jarillo,Marcus,Xu2008,Danneau,Lee,APL2011}, Cr/graphene \cite{NovoselovScience2004,KimNature2005,Han2007,APL2011}, Pd/graphene \cite{Dai2010,APL2011,Avouris2011}, \emph{etc.}, where covalent bonds are formed between the metal and carbon atoms. A strong electronic hybridization takes place so that the linear electronic dispersion disappears \cite{Kelly1,Kelly2}.

Almost all graphene junctions require a wetting layer for the proper stability of the graphene/metal interface, but the impact of such contacts has not been
analyzed by theory in a charge transport setting.
Here we study the transport properties of graphene junctions with Ti contacts with state-of-the-art methods that properly include the effects of tunneling, quantum interference, and contact scattering within the same framework.
 We include the hybridization of graphene $\pi$-electrons with metal $s$-, $p$-, and $d$-electrons that is missing in previous modeling. The potential profile across the junction at zero bias and the conductance as a function of energy are evaluated as a function of $L$, the separation between the metal leads, and $W$, the width of the junctions. The potential variation across the junction under equilibrium conditions is found to be three times bigger than previous estimates based on work-function differences \cite{Kelly2}.

The calculated conductance $G$ provides a measure of the electronic transparency of the junctions. We find that wetting contacts introduce considerable scattering and that the resulting junction transparency exhibits strong energy-dependent fluctuations. We show that  Fabry-Perot oscillations \cite{Dai2010} between the two contacts alone cannot account for that energy-dependence.  This suggests that already the transmission probability of the crystalline contacts is non-uniform. We characterize this non-uniformity first by an analysis of the Fano factor $\mathcal{F}$ of current fluctuations of our junctions. We find   agreement with experiment \cite{Marcus,Danneau} and earlier theory \cite{Beenakker} based on perfect contacts at the Dirac point. The latter observation is  surprising at first sight for two reasons: first,  one expects that contact scattering enhances the Fano factor and second, the agreement  holds up to energies unexpectedly far from the Dirac point. Further statistical analysis reveals that  even the probability distribution of junction transmission eigenvalues agrees with that of a graphene junction with perfect contacts at the Dirac point   \cite{Beenakker} and thus with  the transmission distribution of a diffusive wire \cite{Beenakker,Nazarov}:  transport through the studied junction with wetting metal contacts is pseuso-diffusive to our statistical accuracy within an energy window of width $1 {\rm eV}$ around the Dirac point. We attribute this surprising result to an interplay between the identified contact scattering and the opening of transport channels at finite energy. Our results shed light on the complex transport behavior of graphene junctions with wetting leads and pave the way for a realistic design of potential electronic devices.

\begin{figure}[tb]
\includegraphics[width=0.48\textwidth]{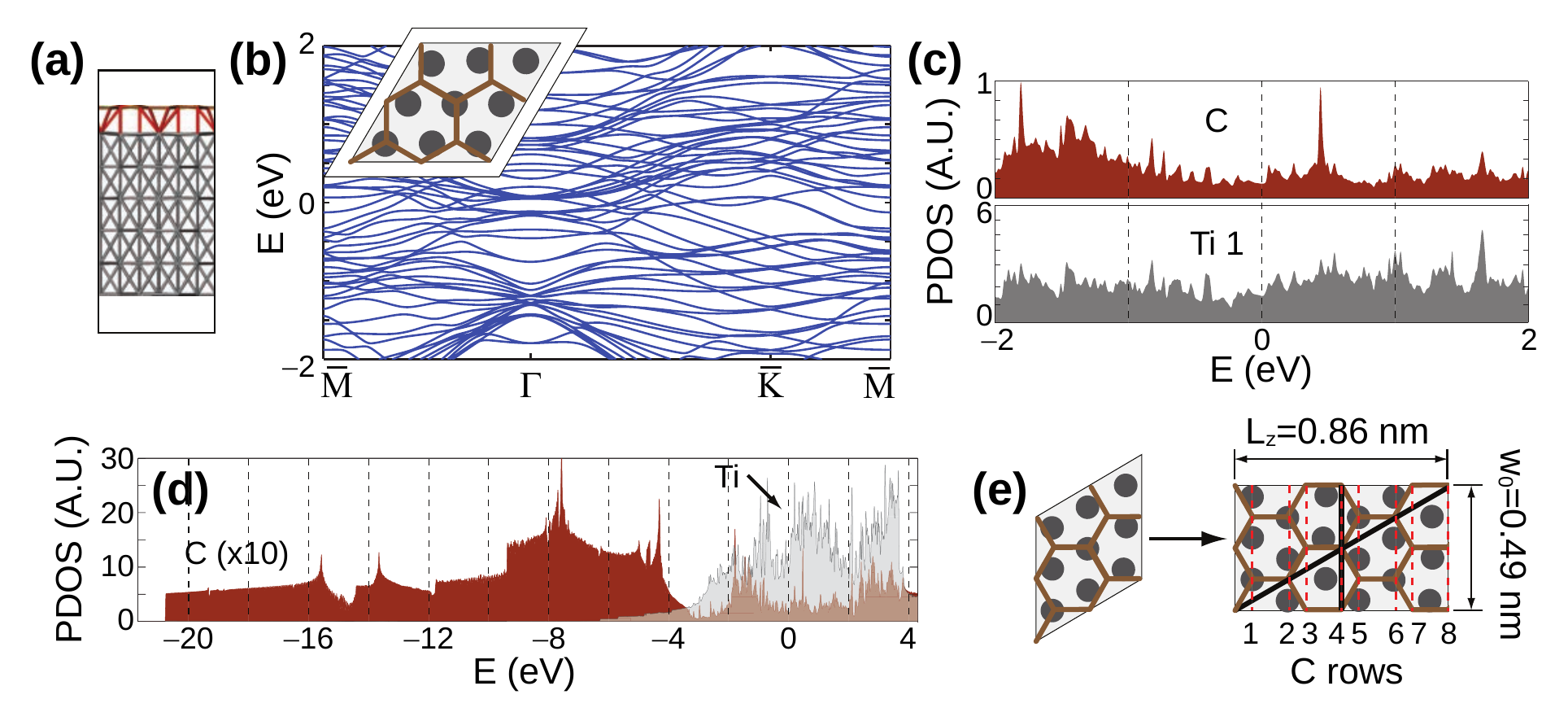}
 \caption{Electronic structure of the lead consisting of graphene and six Ti monolaylers (MLs): (a) Side view of the unit cell. The uppermost layer is graphene, and the Ti/graphene bonds are highlighted in red. (b) The band structure lacks the characteristic linear dispersion of isolated graphene around the $\bar{K}$-point. Since we employ a 2$\times$2 graphene supercell, the original K-point in graphene is folded onto $\bar{K}$. Inset: Top view of the system. (c) Projected density of states (PDOS) on the graphene layer and on the first Ti layer. The energy zero corresponds to the Fermi energy $E_F$ in the lead. (d) PDOS is shown for a larger energy range. (e) Geometry of the rectangular unit cell showing eight C rows  per cell (highlighted by red dashed lines) along the transport direction.}\label{fig:figure1}
\end{figure}

We show in Fig.~\ref{fig:figure1}(a) the atomic configurations for the lead region consisting of six Ti monolayers (MLs) and graphene. The (0001) surface of Ti is in contact with graphene with an average separation of 0.206 nm between the topmost Ti layer and C atoms \cite{Kelly2}. The Ti-C covalent bonds are highlighted in red. We display the corresponding band structures in Fig.~\ref{fig:figure1}(b).
An important observation is that \emph{the linear-dispersion feature of graphene at the $\bar{K}$-point is lost when contacted by titanium}, as generically expected when metals form covalent bonds with graphene \cite{Kelly1,Kelly2}. The hybridization between C and Ti orbitals can be illustrated by the projected density of states (PDOS) on C and on the first Ti layer near the Fermi level. As shown in Fig.~\ref{fig:figure1}(c), many common features appear in the two projections. The PDOS for leads with 6 and 10 Ti MLs are very similar (see Supporting Information), hence in the following we report conductance calculations for the configuration where the leads are formed of graphene and 6 Ti MLs.

The PDOS over a larger energy range is plotted in Fig.~\ref{fig:figure1}(d). While the PDOS associated with the Ti valence electrons (shown in gray) starts at energies of about $-6$ eV, the PDOS for graphene's valence electrons (shown in brown) has contributions from energies as low as $-21$ eV. The carbon PDOS below $-6$ eV mainly comes from the $\sigma$ bonds, which will not be affected by the interaction with Ti except for an energy shift with respect to the Fermi level induced by the presence of the metal \cite{Stadler}. Graphene is heavily $n$-doped by the Ti metal \cite{Kelly2}; we estimate a doping level of 0.1 excess electrons per C atom based on the Voronoi charge analysis \cite{SIESTA2}. (For isolated graphene this charge transfer would have corresponded to a new Fermi level at over 1 eV above the Dirac point $E_D$.)
We display the top view of the unit cell for the contact region in Fig.~\ref{fig:figure1}(e).

\begin{figure}[tb]
\includegraphics[width=0.48\textwidth]{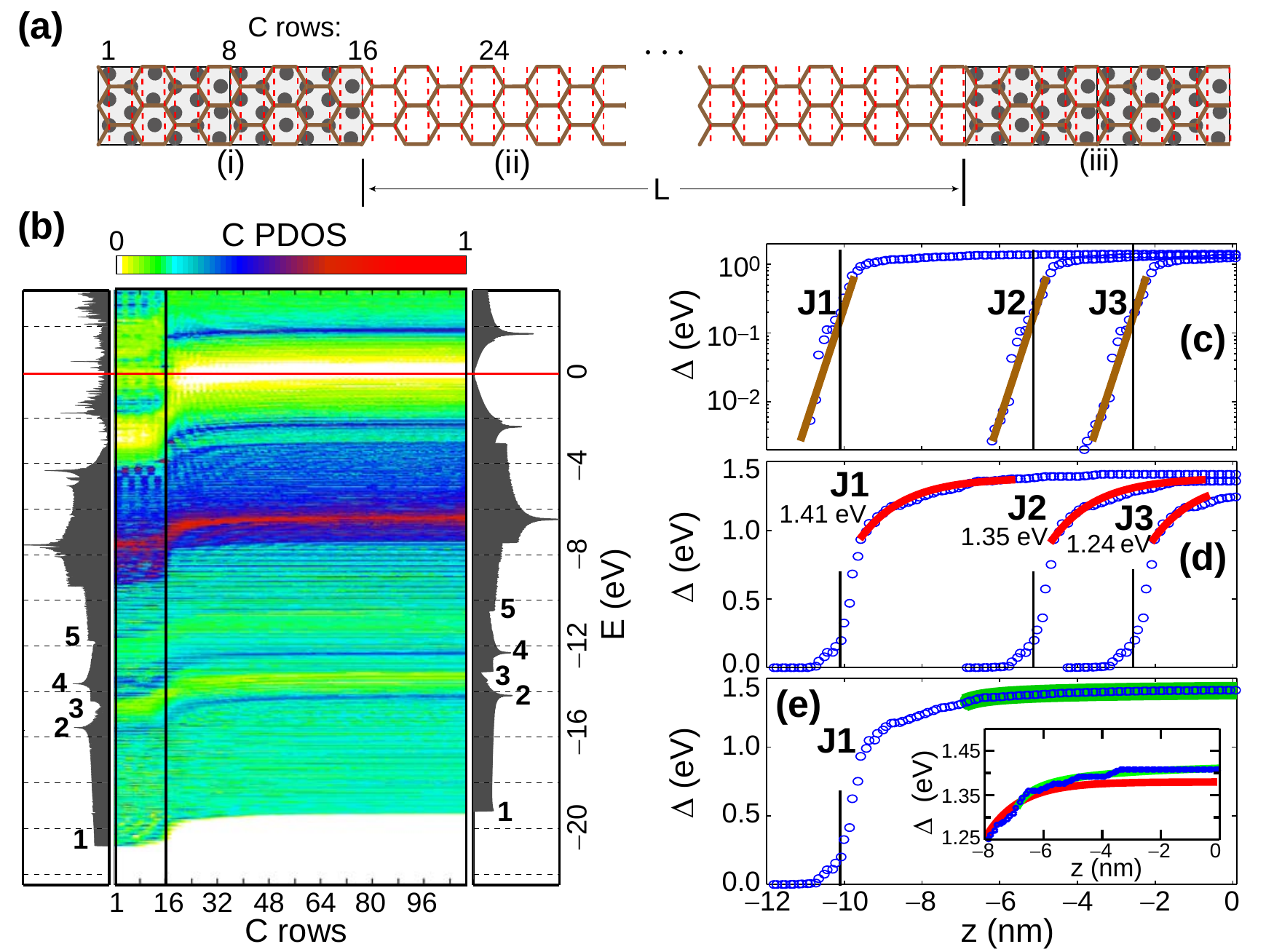}
\caption{Potential profile across the junctions with wetting metal contacts: (a) Schematic top view of the junction; periodic boundary conditions are imposed along the transverse (vertical) direction. (b) PDOS on C atoms across one half of the junction {\bf J1} ($L=20.16$ nm) from density-functional calculations. The solid horizontal line at $E=0$ indicates the Fermi level. We show in gray the PDOS for C located at the first row and half-way between the leads on the left and right, respectively.
(c-e) Potential profiles along the three junctions with lengths of 20.16 nm ({\bf J1}), 10.30 nm ({\bf J2}), and 5.15 nm ({\bf J3}), respectively. The zero coordinate is at the center of the junction. Three regions on this profile can be identified: two exponential variations in (c) and (d), identical for all junctions, and (e) a $z^{-1/2}$ power-law variation (green) for the longest junction {\bf J1}.
}\label{fig:figure2}
\end{figure}

For the conductance calculations we consider a junction consisting of three generic segments shown in Fig.~\ref{fig:figure2}(a):
(i) two lead unit cells [Fig.~\ref{fig:figure1}(e)] to the left, (ii) freestanding graphene over a length $L$, and (iii) two additional lead unit cells to the right. We perform transport calculations using the equilibrium Green's function method with the SMEAGOL package \cite{smeagol1,SIESTA2}. The PBE approximation \cite{PBE} to the exchange-correlation functional and the Troulier-Martins norm-conserving pseudopotentials \cite{TM} are employed, together with an equivalent 400 Ry energy cutoff and the DZP basis sets. We perform structural relaxation until forces are smaller than 0.04 eV/\AA. The complete junctions contain from 232 atoms and 3160 numerical atomic orbitals (NAOs) to 584 atoms and 7880 NAOs. All conductance plots have a 2.5 meV energy resolution. We assume that the leads have a crystalline structure. In order for our large-scale calculations to be feasible, we have imposed a 3.7\% lateral compression for the Ti atoms in the construction of the supercells for the combined graphene/Ti system.

 We have performed calculations for the following values of the separation $L$ between the leads: $20.16$ nm (junction {\bf J1}), $10.30$ nm (junction {\bf J2}), and $5.15$ nm (junction {\bf J3}). In all transport calculations described below, an effective ribbon width is determined by $W = n_k w_0$ \cite{BarrazaPRL2010,Khomyakov2010,Titov}, where $w_0$ is the size of the unit cell in the transverse ($x$) direction shown in Fig.~\ref{fig:figure1}(e), and $n_k$ is the number of $k$-points along this direction used in calculations.

The potential profile for carbon atoms near the metal contact is a property of physical interest, and measurements with a sub-micron resolution have been reported experimentally \cite{Lee}. For graphene junctions with non-wetting metal leads this profile can be directly obtained from the variation of $E_D$ across the junctions: $\Delta (z)=E_D(z)$ \cite{Kelly1,BarrazaPRL2010}. 
For leads with wetting metals as in the present case $E_D$ is not defined at the leads, yet we can still obtain the potential profile in equilibrium from the PDOS.
 Presently, the understanding is that the magnitude of the potential drop across the junction is given by the work function change of graphene on the chemisorbed metal.\cite{Kelly2} For Ti, this yields $\Delta=0.31$ eV, a value much lower than that obtained for physisorbed metals on equilibrium configurations,\cite{Kelly1,Kelly2} where the metal is further away from graphene and not forming bonds (for example, when the physisorbed metal is Al,\cite{Kelly1,Kelly2,BarrazaPRL2010} $\Delta\sim 0.6$ eV). This is a counter-intuitive result, given that charge transfer is about ten times larger when graphene is chemisorbed. Indeed, we recorded 0.1 electrons per C atom while only 0.002 electrons are transferred from Al in the physisorbed case.\cite{Kelly1} One would expect that a larger transfer translates into a larger potential energy shift than that given from the work function difference. We next demonstrate that this is indeed the case.

 We show in Fig.~\ref{fig:figure2}(b) the PDOS of C atoms across junction {\bf J1} up to the mid-point between the leads. The vertical line at C row 16 indicates the edge of the left lead [c.f.~Fig.~\ref{fig:figure2}(a)].
To illustrate the evolution of the PDOS of C atoms versus position
we add the PDOS of C atoms from the lead [Fig.~\ref{fig:figure1}(d)] to the left and the PDOS for pristine graphene to the right of Fig.~\ref{fig:figure2}(b). A number of features remain robust along the junction. We identify five of them in Fig.~\ref{fig:figure2}(b): the lowermost band edge ({1}); two [({2}) and ({4})] van Hove singularities (in dark blue) and a dip between them [({3}) in green]; and a band edge ({5}). Given that the features bend toward negative energies at the lead, the doping of graphene from the wetting metal is $n$-type \cite{Kelly2}, consistent with our charge analysis.

\begin{figure}[tb]
\includegraphics[width=0.48\textwidth]{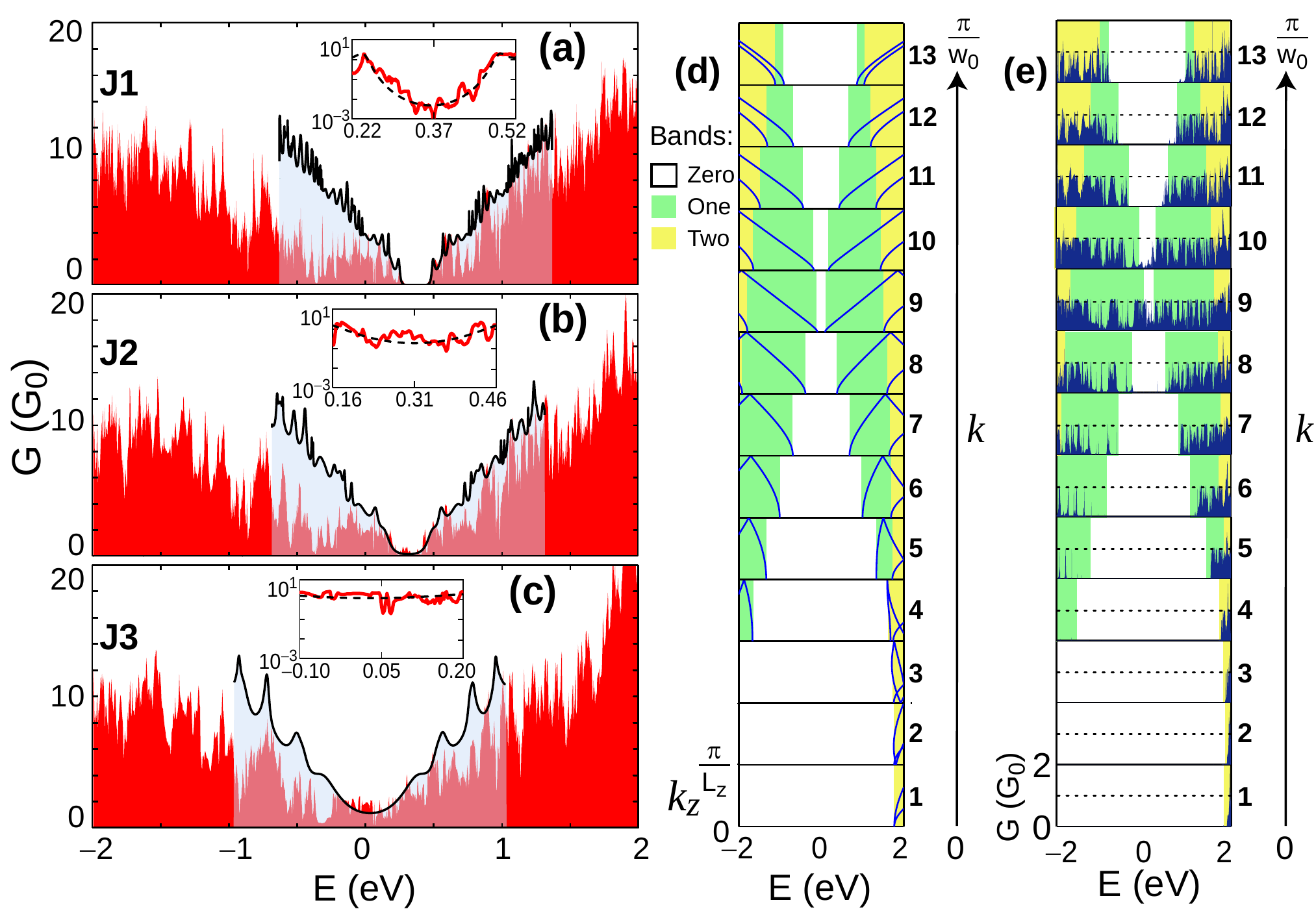}
\caption{Conductance $G$ for graphene junctions with wetting contacts (W=12.38 nm): (a-c) $G$ \emph{versus} $E$. The blue structure corresponds to $G_g$, the conductance through 100\%-transparent graphene. The insets show $G$ (in red) and $G_g$ (dashed line) on a smaller energy range. (d) Band structures at individual transverse $k$-points for a pristine graphene junction. The number of available graphene states are indicated by different colors: 0 in white, 1 in green, or 2 in yellow. (e) Contributions to $G$ from individual transverse $k$-points for junction {\bf J3}.
}\label{fig:figure3}
\end{figure}

Following these PDOS features, we can determine the potential profile variation along the junction $\Delta(z)$, shown in~Fig.~\ref{fig:figure2}(c)-(e).  The edge of the lead is indicated by black vertical lines. Plotted in a logarithmic scale, Fig.~\ref{fig:figure2}(c) indicates that $\Delta(z)$ follows an exponential decay (brown lines of an identical slope) starting from positions well inside the leads. The red curves in Fig.~\ref{fig:figure2}(d) display a exponential variation that is also independent of the junction length \cite{BarrazaPRL2010}.
For short junctions with $L \lesssim 10$ nm a combination of these two exponential curves is sufficient to describe $\Delta(z)$ across the junction \cite{BarrazaPRL2010,Cayssol,Titov}.

The exponential decay flattens out and is unable to capture the behavior towards the middle for the long junction {\bf J1} [see the inset of Fig.~\ref{fig:figure2}(e)]. Instead, a $z^{-1/2}$ power-law \cite{Khomyakov2010} fit shown in green describes the data well. (The parameters and functional forms of all fitting curves are provided in the Supporting Information.) The full variation of $\Delta$ due to the band bending across the junction is 1.41 eV for junction {\bf J1}, 1.35 eV for junction {\bf J2}, and 1.24 eV for junction {\bf J3}, respectively. Importantly, $\Delta$ is substantially larger than the work function change ($|\Delta \Phi_G|=0.31$ eV) for graphene on Ti \cite{Kelly2}. For the shorter junctions there is still residual doping within the freestanding section, yielding a smaller $\Delta$ amplitude.

We have carried out conductance calculations for junctions with an effective width of $W=n_k w_0=12.38$ nm \cite{BarrazaPRL2010} using $n_k=25$. In doing so we neglect boundary effects of the order of $G_0$ ($G_0=2e^2/h$).
Explicitly, we have:
\begin{equation}
G(E)=\sum_{i=1}^{25}G_0T(E,k_i),
\end{equation}
with $T(E,k_i)$ the transmission function at energy $E$ and transverse $k-$point $k_i$.

  The energy-dependent conductance $G$ as a function of $L$ is shown in Fig.~\ref{fig:figure3}(a)-(c). In all conductance plots reported in this Letter $E=0$ corresponds to the chemical potential of the (semi-infinite) leads. Our boundary conditions induce a gap that is inversely proportional to $W$ \cite{PRB96}. The midgap energy $E_c$ is centered at +0.37 eV for junction {\bf J1}, which is, as expected, very close to the change in work function $|\Delta \Phi_G|=0.31$ eV found for graphene on Ti \cite{Kelly2}. In addition, we show in the insets of ~Fig.~\ref{fig:figure3}(a)-(c) and in \ref{table:t1} that $G$ increases exponentially within the energy gap as $L$ decreases, resulting from   charge tunneling between the leads.

A large number of spikes exist in the calculated conductance $G$ shown in ~Fig.~\ref{fig:figure3}(a)-(c), which are not present for junctions with non-wetting contacts \cite{Huard,Beenakker,Cayssol,Golizadeh,Dolfus,BarrazaPRL2010,Venugopal,Nouchi,Avouris2011,Titov}.
 In order to analyze this rather complex conductance pattern, we first display in Fig.~\ref{fig:figure3}(d) the band structures for an isolated sheet of graphene with an effective width $W$ = 12.38 nm at each transverse $k$-point ($k_x$). The number of  bands available for charge transmission (zero, one, or two, as indicated by white, green, and yellow regions in the subplots) determines the maximum conductance contribution of each $k$-point for long junctions, where tunneling is negligible (zero, $G_0$, or $2 G_0$, respectively).
The actual conductance contributions from each of the transverse $k$-points for the shortest junction {\bf J3} ($L$ = 5.15 nm) are displayed in Fig.~\ref{fig:figure3}(e). The height of the conductance peaks in Fig.~\ref{fig:figure3}(e) is indeed consistent with the analysis of Fig.~\ref{fig:figure3}(d), but typically   smaller than the upper bound set by the number of transport channels. This indicates that the contacts in our junctions introduce extra electron scattering, reducing the transmission probability through conductance channels to below 1.   Only at certain energies does the conductance reach its maximum, resulting in the conductance spikes of ~Fig.~\ref{fig:figure3}(a)-(c) that have the appearance of   resonances. Similar results are also obtained for junction {\bf J2} (see Supporting Information).

In order to more accurately quantify the amount of contact scattering   we next compare our junction conductance with the conductance  $G_g$ \cite{Beenakker} of a graphene ribbon with the same dimensions, but fully transparent contacts. In $G_g$ we account not only for propagating graphene modes, but also for tunneling through evanescent  ones.  We plot $G_g$ in Fig.~\ref{fig:figure3}  (in black)  at energies where the theory of Ref.\  \cite{Beenakker} applies.    To account for the energy offset due to the potential $\Delta(z)$ in our junctions, $G_g$ is shifted in energy appropriately. As expected, our junction conductance $G$ is   smaller than its theoretical maximum $G_g$ at almost all energies. Only rarely does  $G$ exceed $G_g$. This excess conductance is predominantly found in our shortest junctions, which suggests that it is due to contact effects that effectively reduce the junction lengths [mainly the induced potential  $\Delta(z)$]. The  junction transparencies, $T_l$ and $T_r$, defined as $G/G_g$ and averaged over energy windows of width $0.25\,{\rm eV}$ about $0.4\,{\rm eV}$ below and above the midgap energy $E_c$, respectively,  are listed in \ref{table:t1}. The transparency $T_r$ is similar to what has been reported experimentally \cite{Jarillo}. $T_l$ is significantly smaller, making $G$ asymmetric. We conclude that  the transparency of graphene junctions is drastically reduced by   wetting contacts. These results represent the first theoretical evaluation of the electronic transparency of wetting graphene contacts.

\begin{table}
\caption{Transparency $T=G/G_g$ for junctions with effective width $W=12.38$ nm averaged over the energy intervals [$E_c-$0.40 eV, $E_c-$0.15 eV] ($T_l$), and [$E_c$+0.15 eV, $E_c$+0.40 eV]  ($T_r$). }
\label{table:t1}
\begin{center}
\begin{tabular}{c c c c c c}
\hline
Junction{~}&{~}$L$~(nm){~}& $E_c$ (eV)&$T_l$&$T_r$\\
\hline
{\bf J1} &20.16 &+0.37      &0.55     &0.74\\
{\bf J2} &10.30 &+0.31      &0.63    &0.75\\
{\bf J3} &5.15  &+0.03      &0.55    &0.77\\
\hline
\end{tabular}
\end{center}
\end{table}

\begin{figure}[tb]
\includegraphics[width=0.48\textwidth]{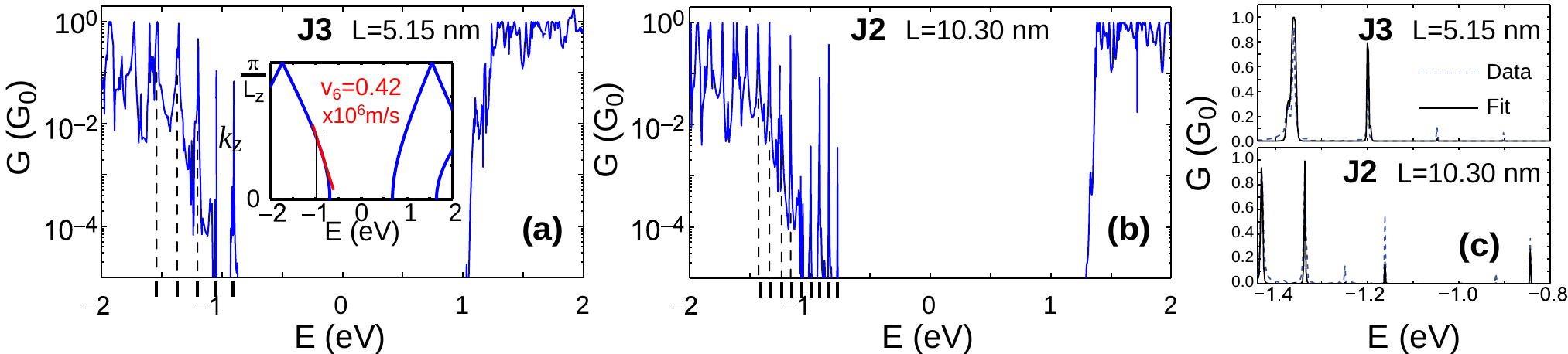}
\caption{Evidence of Fabry-Perot oscillations:
(a) Contribution from $k$-point 6 to the conductance, $G_6(E)$, for junction {\bf J3} ($L=5.15$ nm). Inset: band structure for graphene from Fig.~\ref{fig:figure3}(d) at the sixth transverse $k$-point.
(b) $G_6(E)$ for {\bf J2} ($L=10.30$ nm). Peak spacings are halved when $L$ is doubled (see vertical ticks). (c) Width of the Fabry-Perot oscillations from select data (dashed lines) and theory from Reference 24 (solid line).
}\label{fig:figure4}
\end{figure}

The observed conductance reduction shows that  wetting contacts  introduce electron scattering. At first sight one may therefore attribute the observed conductance spikes to Fabry-Perot (FP) resonances between the two scattering contacts   \cite{Cayssol,Titov}. Such resonances are indeed discernible in the contributions to $G$ from individual $k$-points. For instance, we show the conductance contribution from  transverse ($k_x$) $k$-point 6 in Fig.~\ref{fig:figure4}(a). Plotted on a logarithmic scale, the spectrum clearly displays a series of distinct   peaks with irregular heights, but uniform spacing that is consistent with what is expected for FP oscillations:
The phase difference between two adjacent contributions to the FP interference is $\delta = 2 k_z L$, where $k_z$ is the $k-$vector
along the transport direction [c.f., ~Fig.~\ref{fig:figure1}(e) and Fig.~\ref{fig:figure3}(d)]. Constructive interference occurs when $\delta$ is an integer multiple of $2\pi$. In the inset of Fig.~\ref{fig:figure4}(a), one sees that for the sixth transverse $k-$point ($k_x$) there is an extended energy region where $k_z$ is proportional to $E$ with an effective velocity $v_6$. Therefore, the FP peak separation $\Delta E$ is expected to be $\Delta E(k_6)=\hbar v_6 \pi/L$. For  $L$ = 5.15 nm, we have $\Delta E(k_6)$ = 0.17 eV. This is in excellent agreement with the separation of the peaks below $-1$ eV in the conductance from $k$-point 6 ($G_6$) shown in Fig.~\ref{fig:figure4}(a) and highlighted by the vertical ticks. The energy spacing $\Delta E$ is halved when the junction length $L$ is doubled, as shown Fig.~\ref{fig:figure4}(b), representing additional evidence for FP oscillations.
Additional evidence of Fabry-Perot oscillations is displayed in Fig.~\ref{fig:figure4}(c), where the raw data agrees with $T_{FP}(E)$ in Ref.~\cite{PRLsuggested}. The separation between peaks remains robust as the atomistic structure of the lead is modified (see Supporting Information). However, FP oscillations between simple, energy-independent scatterers clearly cannot account for all features in our conductance traces, in particular not the peak height variations. For energies above the gap  it becomes   difficult to even identify any FP oscillations.

\begin{figure}[tb]
\includegraphics[width=0.48\textwidth]{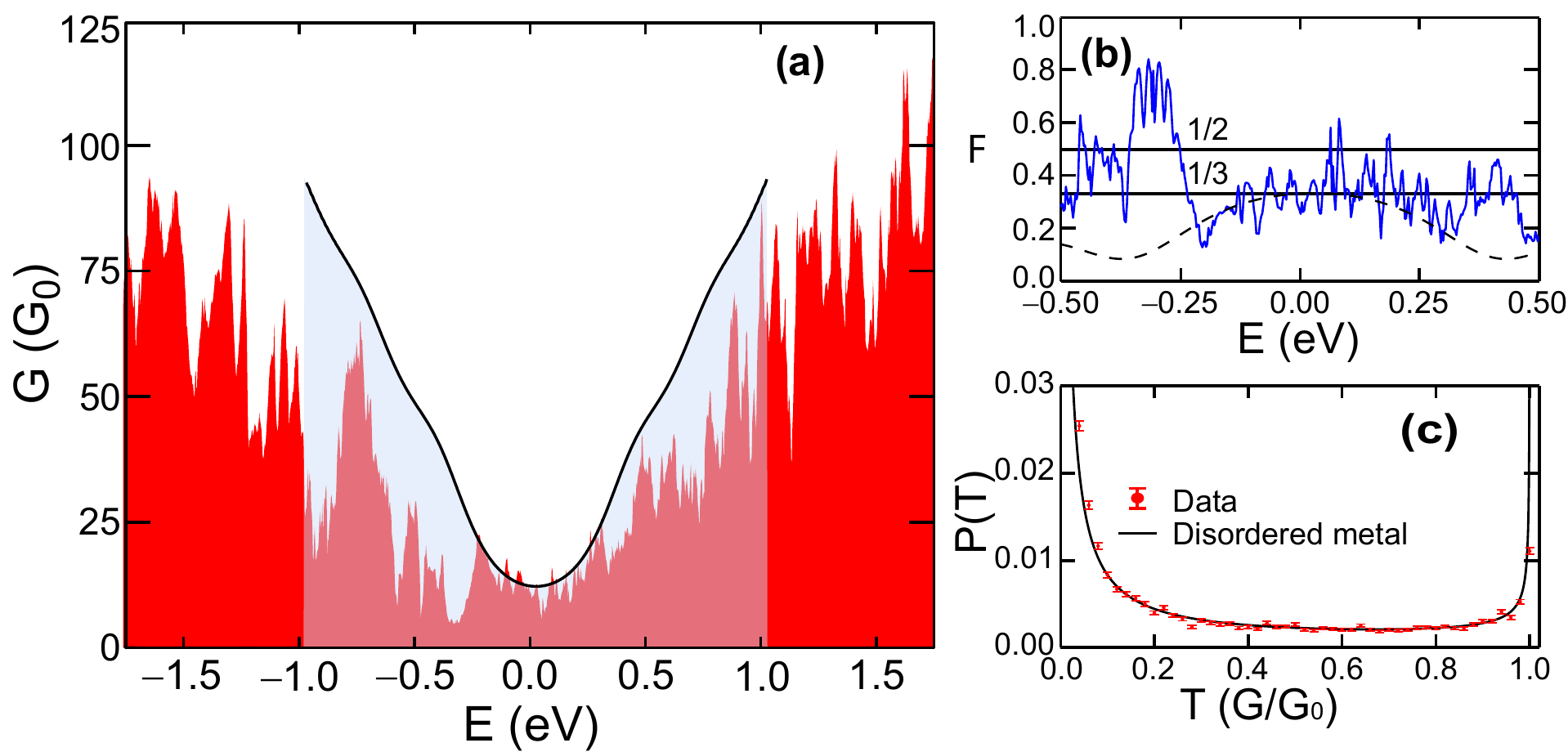}
\caption{Charge transport through a wide junction: (a) Conductance $G$ \emph{versus} energy for a junction with W=99.08 nm and L=5.15 nm.
 The blue pattern $G_g$ is the prediction from the model in Ref.~\cite{Beenakker}.
 (b) Fano factor $\mathcal{F}$ for this junction; the dashed line is a prediction based on the model from Ref.~\cite{Beenakker}. (c) Distribution of transmission eigenvalues. The solid line is the prediction for transmission through a diffusive wire.}
\label{fig:figure5}
\end{figure}
To further characterize the identified fluctuations of the contact transparencies we next compute the shot noise of the studied junctions. Involving the second moment of the transmission probability fluctuations, the shot noise gives statistical information independent of that from the conductance, which is determined by the average transmission. Besides, the shot noise in Ti-contacted graphene junctions has been examined experimentally \cite{Danneau,Marcus}. In order to be in the limit where theory for perfect contacts \cite{Beenakker} predicts universal results we do this analysis for a wider junction. In Fig.~\ref{fig:figure5}(a) we show $G$ for junction {\bf J3} ($L=5.15$ nm) with $n_k=200$ and $W=99.08$ nm.

We find   a Fano factor $\mathcal{F}$ of the shot noise in  {\bf J3}  that  agrees with experiment \cite{Danneau,Marcus} and that  over a wide energy range is surprisingly close to that for  junctions with perfect contacts at their Dirac point:    $\mathcal{F}\approx1/3$   for $-0.5$ eV$<E<0.5$ eV [c.f. Fig.~\ref{fig:figure5}(b)]. Only in a relatively narrow energy interval around $E = -0.3$ eV do we find a substantial increase of $\mathcal{F}$ above $1/3$.
This result is surprising at first sight as additional scattering generically introduces extra shot noise.

Comparison with the shot noise expected for a junction with ideal contacts [c.f.~Fig.~\ref{fig:figure5}(b); black, dashed] reveals one reason for this observation: the Fano factor for a corresponding junction with ideal contacts is suppressed below $\mathcal{F}\approx1/3$ at energies $ |E-E_c|\gtrsim 0.1$ eV because some of the modes that are evanescent at $E=E_c$ become propagating.  At those energies the shot noise enhancement through contact scattering is counteracted by its suppression through the opening of transport modes. Both effects apparently conspire to keep the Fano factor approximately constant at the value  $\mathcal{F}\approx1/3$.  We find $\mathcal{F}\approx1/3$, however, also  at $ |E-E_c|\lesssim 0.1$ eV, where this mechanism is absent. This indicates that contact scattering is suppressed at those energies. Indeed, as seen in In Fig.~\ref{fig:figure5}(a), the conductance of the junction in the vicinity of $E=E_c$ is very close to the expectation for perfect contacts, which indicates a  transparency close to unity. This agrees well with the experiment \cite{Danneau}, where a contact transparency $> 0.8$ was measured.

To examine more closely the correspondence of our junctions in the accessed energy window with ideal junctions at the Dirac point, we characterize the statistics of the fluctuations of all transmission eigenvalues $T$ in the energy window $-0.5$ eV$<E<0.5$ eV by   their probability distribution   $P(T)$. For a junction with ideal contacts at its Dirac point (corresponding to our midgap energy $E_c$) that distribution has been found \cite{Beenakker} to equal $P(T)$  of a diffusive wire \cite{Nazarov,Beenakker2003} at energies $|E - E_c| \lesssim 0.1$ eV.  Remarkably, we find that also the $P(T)$ of   junction   {\bf J3} is to our statistical accuracy identical with that of a diffusive wire, as shown in Fig.~\ref{fig:figure5}(c). We conclude that transport through junction  {\bf J3} in the examined energy window $-0.5$ eV$<E<0.5$ eV is pseudo-diffusive to our numerical accuracy, although an ideal junction displays that behavior only at energies  $ |E-E_c|\lesssim 0.1$ eV. As above, we attribute this surprising observation to an interplay between   contact scattering and transport enhancement as the electron energies depart from the Dirac point.

In summary, we have performed state-or-the-art charge transport calculations for graphene junctions with Ti contacts using Green's functions and density functional theory. Although the potential profile across the junction follows a simple exponential or power-law behavior at different sections, the calculated energy-dependent conductance exhibits strong fluctuations due to contact scattering. We have reported a quantitative estimate of the reduced transparency of these junctions and have analyzed its fluctuations statistically. We have shown that these fluctuations cannot be attributed exclusively to Fabry-Perot oscillations, but that they must be due to energy- and transverse k-point-specific contact scattering.  The statistical distribution of transmission eigenvalues matches to numerical precision that of   a diffusive wire. We conclude that transport through the studied junction is pseudo-diffusive to our statistical accuracy. Accordingly, the shot noise through the studied junctions has  a Fano factor close to 1/3, in agreement with experiment.
 The results presented in this Letter represent a vast improvement over the previous theoretical modeling of transport through graphene junctions, indicating the relevance of the electronic structure at the metal/graphene interfaces for graphene devices.

We thank L. Xian, P. Thibado, K. Park, and M. Kuroda for helpful discussions. S. B.-L. and M. Y. C. acknowledge the support by the US Department of Energy, Office of Basic Energy Sciences, Division of Materials Sciences and Engineering under Award No. DEFG02-97ER45632. M. K. is supported by the National Science Foundation (DMR-10-55799). We thank the support within the Georgia Tech MRSEC, funded by the National Science Foundation (DMR-08-20382) and computer support from Teragrid (TG-PHY090002, NCSA's Ember).\\

\providecommand*\mcitethebibliography{\thebibliography}
\csname @ifundefined\endcsname{endmcitethebibliography}
  {\let\endmcitethebibliography\endthebibliography}{}

\end{document}